\documentclass[sigconf, screen]{acmart}
\usepackage[ruled,vlined,linesnumbered]{algorithm2e}
\usepackage{algpseudocode}
\usepackage{color}
\usepackage{multirow}
\usepackage{amsfonts}
\usepackage{xspace}
\usepackage{colortbl}
\usepackage{makecell}
\usepackage{microtype}
\usepackage{enumitem}
\usepackage{pifont}

\usepackage{amsthm, amssymb}
\theoremstyle{definition}

\newcommand{\tool}{\textit{ContrastRepair}\xspace}

\usepackage{amsmath}

\AtBeginDocument{%
  \providecommand\BibTeX{{%
    \normalfont B\kern-0.5em{\scshape i\kern-0.25em b}\kern-0.8em\TeX}}}

\renewcommand\footnotetextcopyrightpermission[1]{}
\begin{document}

\title{\tool: Enhancing Conversation-Based Automated Program Repair via Contrastive Test Case Pairs}

\author{Jiaolong Kong}
\affiliation{
  \institution{Singapore Management University}
  \country{Singapore}
}

\author{Mingfei Cheng}
\affiliation{
  \institution{Singapore Management University}
  \country{Singapore}
}

\author{Xiaofei Xie}
\affiliation{
  \institution{Singapore Management University}
  \country{Singapore}
}

\author{Shangqing Liu}
\affiliation{
  \institution{Nanyang Technological University}
  \country{Singapore}
}

\author{Xiaoning Du}
\affiliation{
  \institution{Monash University}
  \country{Australia}
}

\author{Qi Guo}
\affiliation{
  \institution{Tianjin University}
  \country{China}
}


\renewcommand{\shortauthors}{Jiaolong Kong, et al.}
\begin{abstract}
Automated Program Repair (APR) aims to automatically generate patches for rectifying software bugs. Recent strides in Large Language Models (LLM), such as ChatGPT, have yielded encouraging outcomes in APR, especially within the conversation-driven APR framework. Nevertheless, the efficacy of conversation-driven APR is contingent on the quality of the feedback information.
In this paper, we propose \tool, a novel conversation-based APR approach that augments conversation-driven APR by providing LLMs with contrastive test pairs. A test pair consists of a failing test and a passing test, which offer contrastive feedback to the LLM. Our key insight is to minimize the difference between the generated passing test and the given failing test, which can better isolate the root causes of bugs. 
By providing informative and specific feedback, \tool enables the LLM to produce effective bug fixes. The implementation of \tool is based on the state-of-the-art LLM, ChatGPT, and it iteratively interacts with ChatGPT until plausible patches are generated. We evaluate \tool on multiple benchmark datasets, including Defects4j, QuixBugs, and HumanEval-Java. The results demonstrate that \tool significantly outperforms existing methods, achieving a new state-of-the-art in program repair. For instance, among Defects4j 1.2 and 2.0, \tool correctly repairs 143 out of all 337 bug cases, while the best-performing baseline fixes 124 bugs.
\end{abstract}

\maketitle



\section{Introduction}
With the increasing complexity of software, the presence of bugs and vulnerabilities has become inevitable. These issues can lead to system failures, security breaches, and a compromised user experience. Manually debugging and fixing these problems is a time-consuming and laborious task, demanding substantial resources and effort from developers. As reported, the annual expenditure on bug finding and fixing amounts to billions of dollars~\cite{britton2013reversible} and developers spend about 50\% of their time in the crucial process of debugging and fixing errors~\cite{latoza2006maintaining, britton2012quantify,zhang2023survey}.
In light of this challenge, Automatic Program Repair (APR) has emerged as a promising solution, offering automated generation of patches to fix bugs.

Automatic Program Repair has been the subject of extensive research in recent years. Traditional APR techniques can be categorized into template-based~\cite{martinez2016astor, liu2019tbar, liu2019avatar}, heuristic-based~\cite{le2016history, le2011genprog, wen2018context}, and constraint-based~\cite{le2017s3, long2015staged, mechtaev2016angelix} methods. However, the effectiveness of these traditional methods is still not satisfactory. For instance, template-based 
APR, considered the state-of-the-art among traditional approaches~\cite{zhang2023survey,xia2023keep,benton2020effectiveness}, requires manually-designed templates, demanding significant human effort and domain knowledge. Consequently, such methods exhibit limited generalization capability, primarily functioning well only on specific types of bugs they were designed for. Recently, machine learning techniques, particularly deep learning-based APR, have gained prominence. These deep learning (DL) models have the advantage of learning diverse patterns of buggy problems from a vast amount of data, surpassing the performance of traditional methods. Nonetheless, DL-based APR techniques still face certain challenges. One significant concern is the reliance on training data; if the training data lacks representation of certain bug types, the model may still struggle to generalize to unseen bugs effectively~\cite{xia2023keep,xia2023conversational}. Moreover, constructing comprehensive bug-fixing datasets for training the DL models requires substantial effort and resources. Furthermore, while DL-based methods show promise in APR, their effectiveness remains limited. These approaches often produce a considerable number of candidate patches, leading to a time-consuming validation process, which presents significant obstacles to the practical implementation of these techniques~\cite{jiang2023impact}.

To overcome the challenges, a promising approach that has gained traction in more recent research is the use of Large Language Models (LLMs). LLMs are trained on vast datasets, such as large-scale code corpora, and have demonstrated superior performance across various tasks~\cite{chen2021evaluating,wang-etal-2021-codet5,xia2023conversational,jiang2023impact,xia2023keep}. The primary factor for their superior performance lies in the remarkable ability to comprehend program semantics. In the context of APR, studies \cite{10.1145/3540250.3549101,xia2023automated, prenner2022can, jiang2023impact, sobania2023analysis} have shown that even without fine-tuning, LLMs exhibit competitive fixing capabilities compared to traditional DL-based APR techniques. The fixing capability can be significantly enhanced by fine-tuning the LLMs on relevant data~\cite{jiang2023impact}.
Previously, most of the works invoke LLMs independently without incorporating conversation. With the advance of dialog-based LLMs, such as ChatGPT, recent conversational-driven APR based on ChatGPT has achieved new state-of-the-art performance~\cite{xia2023keep,xia2023conversational}. The basic idea is to generate patches in a conversational style. At each conversation, the system combines previous incorrect patches with test failure information to prompt the LLM to generate a new patch. This \textit{conversational-driven} APR has demonstrated remarkable performance in guiding the LLM to propose more effective repairs.

This paper primarily focuses on the utilization of LLMs for \textit{conversational-driven} APR. The paramount challenge in applying LLMs for this purpose is the formulation of high-quality prompts, which play a crucial role in guiding the LLM to comprehend programs and subsequently fix bugs. While recent research~\cite{xia2023keep} has shown promising results using LLMs, the proposed approach mainly depends on providing error feedback information for the conversations. Unfortunately, this feedback might not always offer specific and informative prompts for effective repair.




To this end, we propose a method that aims to craft more specific and informative prompts for enhancing the capabilities of LLMs in accurately localizing bugs and generating high-quality fixes. Our key insight lies in that relying solely on negative feedback, i.e., error information from running failing tests, may not always be adequate for LLMs to precisely pinpoint the bugs. We propose the inclusion of positive feedback, derived from successful tests, to supplement the negative feedback.  By providing comparative input pairs that juxtapose the outcomes of these tests, LLMs are more likely to effectively localize the root cause. We propose a novel conversation-driven approach \tool that generates prompts by utilizing both positive and negative feedback.  Given a failing test case and the buggy function, we create a corresponding passing test case that is quite similar to the failed one, forming a contrastive pair that is then forwarded to the LLMs\footnote{While our approach is general, we specifically chose ChatGPT in this work due to its state-of-the-art performance in handling conversational-style interactions.}. By providing such pair-wise information, LLMs receive more specific and informative cues, allowing them to better pinpoint the root cause of the bug and generate accurate fixes. Nevertheless, a challenge lies in selecting the appropriate passing test case (from a considerable number of passing tests) to pair with the failing test case. Our basic idea is to construct the test pair (i.e., the failing test and the passing test) by minimizing their difference, such that the difference (which makes one test fail and another one pass) can effectively isolate and pinpoint the failure causes. 
Specifically, we present a method designed to generate a passing test case by making minimal modifications to a failing test case. If there already exists a collection of passing test cases, we then select the most similar one from the set. In addition, some bugs may have strong dependencies on other functions, which is also an important context for the repair. We incorporate more contextual information by selecting relevant functions from the traceback of the failing test case.
Finally, the prompt provided to the LLM consists of the buggy function, the dependent functions, the test pair, and the traceback information from the failing test case. The LLM then outputs a patch function. If the patch is deemed incorrect, we adopt a similar approach to iteratively generate the next patches. This iterative process continues until plausible patches, i.e., those capable of passing all the tests, are produced, or the repair budget is exhausted. 



We evaluated \tool on three distinct datasets: Defects4J, QuixBugs, and HumanEval-Java. Specifically, we compared the performance of \tool against state-of-the-art approaches, encompassing both learning-based APR methods and conversation-based APR methods, such as CHATREPAIR \cite{xia2023keep}. The results of our evaluation clearly demonstrate that \tool significantly outperforms the baseline methods. Notably, \tool successfully resolved 360 out of 581 bugs, surpassing the state-of-the-art ChatGPT-based method CHATREPAIR, which could fix 334 bugs. Additionally, our findings reveal that \tool exhibits enhanced efficiency in terms of API calls with an average reduction of 20.91\% compared to CHATREPAIR. 
Furthermore, our ablation study underscored the significance of pair selection and contextual information in augmenting the performance of \tool.


In summary, this paper makes the following contributions
\begin{itemize}[leftmargin=*]
\item 

We present a novel approach for APR that leverages Large Language Models in a conversation-driven manner. Through the integration of both negative and positive feedback, our approach empowers LLMs to provide high-quality fixes.

\item 
 We propose a method to construct and prioritize suitable test pairs comprising both failed and corresponding passing tests. This method ensures that the prompt provided to LLMs is more informative, aiding in precise bug localization and more effective patch generation.

\item
We perform a comprehensive evaluation to evaluate the effectiveness of our tool. The results demonstrate that \tool achieves new state-of-the-art performance in terms of correct bug fixes, increasing by 15.32\% than the best baseline, surpassing existing APR tools.
\end{itemize}

\begin{figure*}[!t]
    \centering
    \includegraphics[width=1\linewidth]{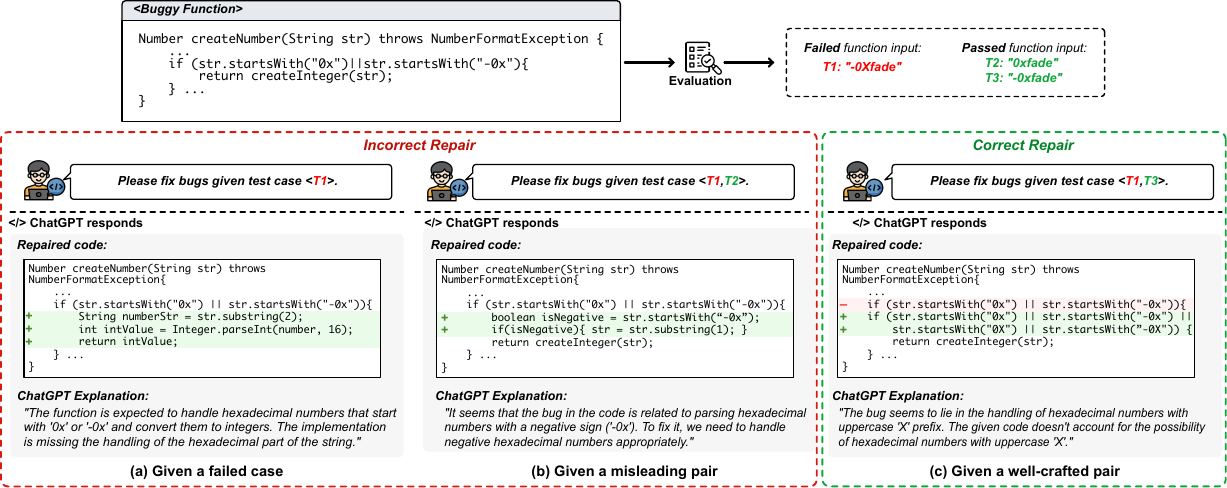}
    \caption{A Motivating Example}
    \label{fig:motivatingexample}
\end{figure*}

\section{Motivation Example}


\label{sec:motivating}
Fig~\ref{fig:motivatingexample} depicts a motivating example (Lang-16 from Defects4j) that comprises a buggy function and a set of test cases designed for evaluating this function. By giving different types of test cases, we show the results of ChatGPT, including fixed code and the explanation about the fixes.
Specifically, \texttt{createNumber} works to parse a given string \texttt{str} and returns its corresponding Number object. 
A subtle bug is present in the function code.
When \texttt{str} represents the string of a hexadecimal integer, the function correctly handles cases where it starts with lowercase ``0x'' or ``-0x''. However, it overlooks the uppercased inputs, leading to incomplete handling of corner cases. Consequently, when evaluating \texttt{createNumber} on three test cases, both T2 and T3 pass, while T1 fails.

When fixing this program, the repairer needs to first locate the buggy line(s) and then correctly address the missed corner case.
Relying on ChatGPT, although it exhibits a good understanding of program semantics, additional guidance on the requested task remains beneficial. 
While the failing test case T1 is supplied to aid in the repair process and narrow down the search space by providing hints on the buggy lines and potential fixes, the impact of failing test cases alone is somewhat limited. 
As a result, more effective guidance is desired to achieve better results. 




On the other hand, when we provide a well-crafted contrastive test pair {<T1, T3>}, such as (``-0Xfade'', ``-0xfade''), it significantly simplifies the task for the repairer to deduce that the characters `X' and `x' may be the problematic elements causing the bug. 
This enables the repairer to accurately pinpoint the buggy lines, i.e., those dealing with problematic elements, and propose a suitable fix.
It shows how contrastive information derived from both positive and negative feedback can offer more precise guidance for program repair. By leveraging such contrastive test pairs, the repair process becomes more efficient and effective.
However, it is crucial to recognize that not every test pair provides valuable guidance for the repair process. In some cases, the guidance from certain comparisons may be indirect or even incorrect. For instance, the pair {<T1, T2>}, i.e., (``-0Xfade'', ``0xfade''), could lead ChatGPT to mistakenly infer that the symbol `-' is the cause of the bug, which could be misleading for the repair. 
Thus, \textit{how to better select and craft contrastive test pairs is of great importance to ensure the effectiveness of the repair guidance}.

\section{Methodology}
\subsection{Overview of \tool}

\begin{figure*}[!t]
    \centering
     \includegraphics[width=1\linewidth]{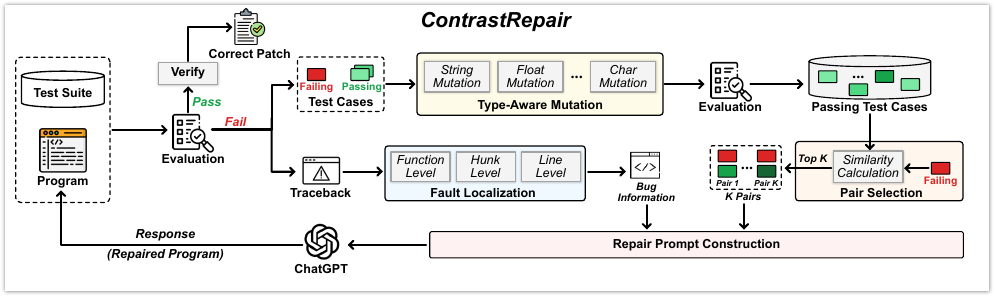}
    \caption{Overview of \tool}
    \label{fig:overview}
\end{figure*}

Fig~\ref{fig:overview} provides an overview of the \tool. Essentially, it involves a conversation process that constructs prompts using contrastive test pairs, feeds these prompts to an LLM such as ChatGPT, and receives responses that lead to the repaired code.

Specifically, at each iteration of the conversation, \tool takes as input a program to be validated, along with a test suite consisting of multiple test cases. The initial program is assumed to be buggy code. \tool then evaluates the program with the test suite. If all tests pass successfully, a plausible patch is obtained, which is further investigated by a human. On the other hand, if any test fails ($f$), and triggers the bug in the program, then the corresponding traceback logs ($T_f$) are captured. Additionally, all passing tests ($P$) that the program handles correctly are also collected for further test pair construction.

Next, \tool constructs test pairs that involve the failing test $f$ and other passing tests (e.g., $f'$, $f''$). To ensure these test pairs provide informative contrastive guidance, we propose a similarity-guided selection process. 
Inspired by the concept of Delta Debugging~\cite{zeller2002simplifying}, which aims to isolate failure causes by minimizing the \textit{failing test}, we select a passing test that has minimal difference from the failing test. This enables ChatGPT to isolate failure causes by contrasting their minor differences. In cases where no existing test cases have high similarity with $f$, we introduce a type-aware test mutation technique to generate a satisfactory passing test based on its type. It minimizes the mutation of the failing test case, thereby enhancing the contrastive information that is helpful in isolating the failure causes.

We then follow the best practices~\cite{chat_prompt} to construct prompts for ChatGPT. To collect the dependency information of the bug, we identify the dependent functions from the traceback of the failing test case. The prompt consists of the traceback logs from the failing test case $f$, the constructed test pairs, the dependent functions, the target buggy function, and the description of the requirements. With this prompt, ChatGPT generates the repaired code, which is then validated in the subsequent iteration. The iterative process continues until a plausible patch is identified or the repair budget (e.g., the maximum number of iterations) is reached.


\vspace{-2mm}
\subsection{Constructing Contrastive Test Case Pairs}
A common sense for code debugging is that ``Often people who encounter a bug spend a lot of time investigating which changes to the input file will make the bug go away and which changes will not affect it''~\cite{zeller2002simplifying}. 
The changes that make the bug go away are called \textit{critical changes}. These critical changes hold valuable insights into understanding the root cause of bugs, making them essential for successful bug localization and repair. Therefore, our objective is to identify and provide critical change information to LLMs. 

To facilitate the exposure of the crucial changes to LLMs, we present a method that involves constructing contrastive test case pairs, consisting of a failing test case and a corresponding passing test case. We ensure that the failing test and the passing test are sufficiently similar, which allows LLMs to deduce the critical changes based on their minor differences that lead to the triggering of the bug.

Specifically, to create the contrastive test case pairs, we first construct a set of passing tests $P$. Then we select the suitable passing tests (denoted as $S$) from $P$ to pair with the given failing $f$, ensuring that they exhibit higher similarity with $f$:
$S=\{p| p\in P \wedge \delta(f, p)>\theta\}$.
where $\delta$ represents the similarity measurement between the failing test $f$ and a passing test $p$, and $\theta$ is a predefined threshold.
Two main problems need to be addressed: how to measure the similarity ($\delta$) and how to construct the passing test set ($P$).

\subsubsection{Similarity Measurement.} 
Various similarity metrics can be considered. However, since our primary objective is to expose critical changes regarding the failing test, a string-based similarity metric is more relevant for this task. In this paper, we select the \textit{Damerau-Levenshtein distance}~\cite{damerau1964technique, levenshtein1966binary} as the chosen similarity measure. The Damerau-Levenshtein distance is a metric used to gauge the similarity between two strings by quantifying the minimum number of operations required to transform one string into the other. It accounts for four types of edit operations, i.e., insertion, deletion, substitution, and transposition. A smaller Damerau-Levenshtein distance indicates greater similarity between the two cases, which is more beneficial for LLMs to isolate the root causes. We normalize the similarity score to the range of (0, 1), i.e., $\delta(f, p)=1-d/max(len(f), len(p))$.

Note that the test cases may have different types. To accommodate this, we convert the tests with other types into string representation before calculating their similarity using the Damerau-Levenshtein distance. For object or array types, we recursively convert each element to its corresponding string representation and then calculate the similarity.

\vspace{-2mm}
\subsubsection{Passing Test.} 
To gather the passing test cases, we adopt two strategies: utilizing existing passing tests and generating new passing tests. Many projects, such as Defects4j, include unit tests that are designed to validate the correctness of individual functions. We execute all unit tests and collect the corresponding passing test cases for each function. In cases where certain functions lack unit tests or the existing unit tests have low similarity with the failing test $f$, we propose a type-aware mutation technique to generate new passing test cases. The mutation is based on the failing test $f$ with changes made as minimally as possible.

\textit{Type-aware Mutation.} 
The type-aware mutation comprises two main components: a general string-based mutation and type-specific mutations. Given that the similarity is measured based on string representations, the natural approach is to first convert the failing test $f$ to a string representation. We then perform string-level mutations while constraining the degree of mutations. Finally, we convert the string mutants back to their original types based on a Python lib-javaobj~\cite{javaobj}. Additionally, we employ type-specific mutations to enhance the diversity of the generated test cases. The mutation strategies are listed as follows:

\begin{itemize}[leftmargin=*]
\item \textit{String Mutation}: 1) Random Character Replacement; 2) Random Substring Replacement; 3) Random Character Insertion; 4) Random Character Deletion; 5) Substring Swapping; 6) Case Conversion and 7) Truncation/Extension.

\item \textit{Integer, Double, and Float Mutation}: 1) Random Perturbation, adding or subtracting a small random value; 2) Scaling, multiplied by a scaling factor; 3) Flip Sign, changing from positive to negative or vice versa and 4) Magnitude Perturbation, adding or subtracting a small percentage of the number's magnitude.

\item \textit{Char Mutation}. Replace the original character with a randomly chosen different character (Random Replacement).

\item \textit{Boolean Mutation}. Negate the boolean value (Negation).

\item \textit{Object Mutation}. Element-wise mutation is based on the mutations on primitive types above.

\item \textit{Array, List}: 1) Element-wise Mutation, mutating one or more elements; 2) Element Swapping; 3) Element Insertion; 4) Element Deletion and 5) Element Shuffling.
\end{itemize}

As functions may have multiple parameters, we consider each parameter as an element or attribute of an object. Consequently, we treat the set of parameters as an object-type test case. This allows us to conduct object-level mutation and similarity comparisons.

\textit{Test Oracle.} The collection of passing test cases poses a challenge due to the lack of test oracles, especially for logical bugs. We categorize bugs into two groups: exception bugs, for which the test oracles can be determined based on whether the exception is thrown, and logical bugs, which can only be captured through explicit assertions.
For exception bugs, we employ type-aware mutation to generate passing tests. However, for logical bugs, in this paper, we primarily rely on existing assertion-based test cases to serve as our test oracles for logical bugs. If there is no existing passing test case, we use a single failing test instead of a pair. We leave the integration and exploration of existing solutions for addressing logical bug oracles (e.g., constructing metamorphic relations) as our future work.

\newcommand\mycommfont[1]{\footnotesize\ttfamily\textcolor{blue}{#1}}
\SetCommentSty{mycommfont}
\SetKwInput{KwInput}{Input}                
\SetKwInput{KwOutput}{Output}              

\begin{algorithm}[!t]
\small
\DontPrintSemicolon
  \KwInput{$Func$: the buggy function, $\alpha$: the whole tests, $k$: the number of pairs selected,
  
  $m$: the maximum number of restarting repair attempts, $n$: the maximum number of continuous repair attempts
  }
  \KwOutput{$PFuncs$: the plausible patch(es)}
   $P, F := \texttt{Collect}(Func, \alpha)$; \tcp*{collect passing and failing tests}\label{algo:collect}
   $\phi := \{(f,p)| f\in F \wedge p\in P \wedge \delta(f,p) > \theta\}$; \tcp*{get all similar pairs}\label{algo:getpair}
  $iter_1:=0$;\\ 
    \While{$iter_1 < m$}{\tcp*{\textbf{restart} repair from original function} \label{algo:whilerestart}
        $\rho := \texttt{SelectPair}(\phi, k)$; \tcp*{select $k$ pairs} \label{algo:selectk}
        $tmp := Func$;\\
        $iter_2:=0$;\\
        \While{$iter_2 < n$}{\tcp*{\textbf{continuous} repair from previous patch}\label{algo:whilecontinue}
            $T := \texttt{Execute}(tmp, \rho_f)$; \tcp*{get traceback info} \label{algo:gettraceback}
            $D := \texttt{ExtractDF}(Func, T)$; \tcp*{get dependent functions from $T$} \label{algo:getdependent}
            $prompt: = \texttt{ConstructPrompt}(tmp, \rho, T, D)$;\\ \label{algo:constructprompt}
            $Func' := \texttt{ChatGPT}(prompt)$;\\  \label{algo:invokechatgpt}
             $P', F’ := \texttt{Collect}(Func', \alpha)$; \tcp*{evaluate the patched function} \label{algo:reevaluateandcollect}
            \If{$F'=\emptyset$}{
                $PFuncs:=\texttt{PatchAug}(Func')$ \label{algo:patchaug}\tcp*{generate alternative plausible patches}
                \textbf{return} $PFuncs$;\label{algo:correctpatch} \tcp*{plausible patches}
            }
            $\phi' := \{(f,p)| f\in F' \wedge p\in P' \wedge \delta(f,p) > \theta\}$;\\ \label{algo:phiagain} 
            $\rho := \texttt{SelectPair}(\phi', k)$;\\
            $tmp := Func'$;\tcp*{update $\phi'$, $\rho$ and $tmp$} \label{algo:funcagain}
            $iter_2 = iter_2 + 1$;
        }
         $iter_1 = iter_1 + 1$;
    }
    \Return $Func$;  \tcp*{fail to repair}

\caption{\tool}

\label{algo}
\end{algorithm}

\subsection{Conversation-Driven Repair}
The repair process follows an iterative approach that involves interacting with LLMs to gradually repair the buggy function. Two main strategies can be employed during the repair process:

\begin{itemize}[leftmargin=*]
    \item 
\textit{Continuing from Previous Patch:} In this strategy, at each iteration, we continue the repair process from the previous patch that is still not correct. This approach allows us to build on the previous repair attempts and leverage continuous feedback from LLMs to refine and improve the patches incrementally. However, there is a risk that the repair process may continue in the wrong direction, potentially worsening the patches.
\item 
\textit{Restarting Repair from Original Buggy Function:} Alternatively, at each iteration, we can start the repair process from the original buggy function. This strategy ensures that each repair attempt begins from a clean state, reducing the risk of compounding errors from previous iterations. However, it also loses the continuous feedback that can be valuable for guiding the repair process.
\end{itemize}

To strike a balance between these two strategies, we impose a limit on the depth of continuous repairs. If the function cannot be repaired within the specified maximum depth, we restart the repair process from the original buggy function and select different test pairs. This approach allows us to explore multiple repair paths and increase the chances of generating accurate fixes.

Algorithm~\ref{algo} presents the overall repair process of \tool. The algorithm takes as input the buggy function and a set of test cases $\alpha$ used to evaluate its correctness. The objective is to output a plausible patch if successful. 
\tool collects all failing test cases and passing test cases by running all available tests (Line~\ref{algo:collect}). In cases where the number of passing tests is insufficient or they have low similarities with the failing test, the function \texttt{Collect} includes the type-aware mutation to generate additional passing tests for exception bugs. Then it selects all contrastive test pairs that exhibit sufficient similarity (Line~\ref{algo:getpair}). The repair process involves both restarting the repair (Line~\ref{algo:whilerestart}) and continuous repair (Line~\ref{algo:whilecontinue}). The total number of repair attempts is limited to $m \times n$, where $m$ is the maximum tries for the restarting repair and $n$ is the maximum tries for the continuous repair.

At each restarting iteration, $k$ test pairs are selected for constructing prompts (Line~\ref{algo:selectk}). The \texttt{SelectPair}
function aims to prioritize pairs based on how often a pair is selected before. The pairs that were rarely selected have high priority, which prevents an excessive number of selections of the same pairs. For the selected pairs $\rho$, the algorithm performs continuous repair within $n$ tries. 
{The failure test cases in the $k$ pairs (denoted as $\rho_f$) may or may not be caused by the same root cause. We execute all of them to collect the bug traceback logs and remove the redundant duplicates to reduce the tokens sent to LLMs (Line~\ref{algo:gettraceback})}. In cases where the bug depends on other functions, \tool identifies the dependent functions from the traceback logs, {then merges them together and forms the dependent function collection $D$} (Line~\ref{algo:getdependent}). 
The prompt is constructed (Line~\ref{algo:constructprompt}), including the buggy function $tmp$, the test pairs $\rho$, the traceback $t$, dependent functions $d$, and specific repair requirements. The prompt is then fed to ChatGPT, and the patched function is obtained (Line~\ref{algo:invokechatgpt}). 

The patch is evaluated with all tests (Line~\ref{algo:reevaluateandcollect}). If passed, a plausible patch is generated. Following the work~\cite{xia2023keep}, we generate alternative plausible patches by ChatGPT in case the generated patch is not the correct patch but is very close to it (Line \ref{algo:patchaug}).
Otherwise, the repair process continues from the current patched function and its test pairs (Line~\ref{algo:phiagain}-\ref{algo:funcagain}). If the repair fails within the given budget, the original buggy function is returned.

\begin{figure*}[!t]
    \centering
     \includegraphics[width=1\linewidth]{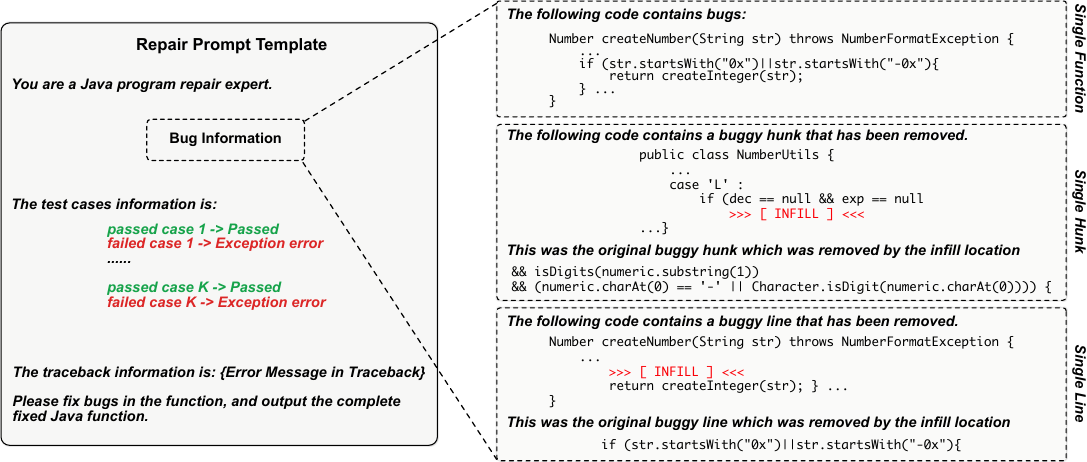}
    \caption{Illustration of Prompt Construction}
    \label{fig:prompt_construction}
\end{figure*}

\textit{Prompt Construction}.
The construction of prompts is a crucial step in effectively guiding ChatGPT for program repair. Following best practices~\cite{guo2023exploring, white2023prompt, chat_prompt}, we design the prompt to include essential elements that provide context, task description, and constraints on the output. We manually examined several alternative prompts using the web version of ChatGPT and selected the best one. 
As is presented in Fig~\ref{fig:prompt_construction}, the prompt begins by setting the context for ChatGPT, introducing it as a ``Java program repair expert.'' Then it includes the code of the buggy function that requires repair, along with the contrastive test pair that was constructed. The traceback information from the failed test is added to the prompt, providing additional context and insights into the bug. If the bug depends on other functions, we include the code about these dependent functions. Specifically, we extract the function names that appear in both the traceback of the error message and the source code of the project. Those that are direct callers or callees of the buggy function are considered dependent functions. If a dependent function depends on other functions, then others are also taken into account unless the length of the prompt hits the maximum limit. The dependent function could provide more information for ChatGPT to understand the buggy program and the root causes of bugs. The prompt concludes with specific requirements, instructing ChatGPT to generate a complete patched function.

\vspace{-2mm}

\section{Evaluation}
In this section, we aim to answer the following research questions:

\begin{itemize}[leftmargin=*]
\item
\textbf{RQ1:} \textit{How effective is \tool compared to state-of-the-art APR techniques?} We evaluate the performance of \tool by comparing it to traditional APR methods, DL-based APR techniques, and the recent conversation-driven APR tools.
\item
\textbf{RQ2:} \textit{How useful is \tool in unknown datasets?} Considering the potential data leak risk caused by LLMs, we evaluate \tool on previously unseen datasets, which were not used during the training of ChatGPT.

\item
\textbf{RQ3:} \textit{How do different hyperparameters affect the repair performance of \tool?} We conduct a study to examine the impact of various hyperparameters on \tool's effectiveness, including the number of test pairs, and the thresholds for continuous repair and restarting repair.

\item
\textbf{RQ4:} \textit{What are the contributions of different components of \tool in improving repair effectiveness?} We aim to understand the usefulness of each component of \tool, including the contrastive test pairs and the dependent functions.

\end{itemize}


\subsection{Setup}


\subsubsection{Configuration}
For the experiments conducted in this paper, we opted to utilize ChatGPT with the gpt-3.5-turbo-0301 model as the pre-trained Large Language Model (LLM) for \tool.
To interact with the ChatGPT service, we employed its API.
In order to increase the potential search space and generate diverse patches, we set the sampling temperature to 1, which is the default setting. The maximum number of continuous repair attempts ($n$) and the maximum number of restarting repair attempts ($m$) were set as 3 and 40. Therefore, the maximum number of conversations per buggy function is 120. Specifically, in the process of querying ChatGPT to repair bugs, there is no timeout setting, when query times are used up, the repair process stops. Once a plausible patch is generated, we take patch augmentation to generate more plausible patches by prompting ChatGPT with already collected plausible ones, and we set query times for this process to 40. Based on an initial study on similarity distribution after mutations (see RQ1), we set the similarity threshold, $\theta$, at 0.5. This decision balances the number of test pairs we can select while aiming to maximize their similarity.
Instead, we generated a set of test cases by mutating the failing test from a small to a large degree or by collecting existing passing tests. For the mutation of the failing test, we randomly generate 1,000 new test cases for validation and selection. And in the process of evaluating newly generated test cases, there is a timeout. Specifically, We set the maximum time as 30 seconds to evaluate each generated test case and verify whether the buggy function can pass it. With this setting, the time cost for each bug in the process of type-aware mutation(the time cost for mutation is really little, which can be omitted) and validation of newly generated test cases can be limited to 25 minutes. Notably, in our approach, test generation is only invoked once before the repair process starts. 


\subsubsection{Benchmark}

We evaluate the effectiveness of \tool by using well-established benchmark datasets, including Defects4j and QuixBugs. Defects4j is a widely studied collection of open-source bugs from 15 different projects, while QuixBugs consists of 40 buggy and fixed versions of classic programming problems in both Python and Java.

To provide fault localization during the repair process, we adopt the approach used in prior research~\cite{xia2023keep,10.1145/3540250.3549101} by providing perfect fault locations derived from ground truth. We evaluate the performance of the tools in three different scenarios based on the granularity of the provided fault locations: at the function level, the statement level, and the hunk level. These scenarios correspond to single-function fixes, single-line fixes, and single-hunk fixes, respectively. It is worth noting that function-level fault localization is more practical and more challenging for repair, given that fine-grained fault localization is often difficult to obtain in real-world situations.

For the Defects4j dataset, we follow the standard practice in prior APR works and divide it into two versions: Defects4j 1.2 and Defects4j 2.0. Defects4j 1.2 contains a total of 391 bugs in 6 different Java projects. Based on the fault localizations provided, they can be categorized as 255 single-function bugs, 154 single-hunk bugs and 80 single-line bugs. Due to the budget constraint for invoking the ChatGPT API, for Defects4j 2.0, we select 82 single-line bugs, which is a commonly used setting in prior APR tools~\cite{xia2023keep,10.1145/3540250.3549101} for easier comparison. Notably, the category of single-line bugs is a subset of the single-function bug category and single-hunk bug category, and the single-hunk category is also a subset of the single-function category.
In addition, we also assess the performance of \tool on the QuixBugs dataset, which consists of both Java and Python bugs, 

Additionally, for our evaluation, we incorporated the HumanEval-Java ~\cite{jiang2023impact} benchmark, which has 163 bug cases and most of them are single-line ones. The release date of HumanEval-Java postdates the data collection used for training GPT-3.5, ensuring that there is no risk of data leakage. Within this benchmark, developers translated Python programs from HumanEval and their corresponding test cases into Java programs and JUnit test cases. Some bugs are introduced into the correct Java programs. Similarly, we also provide function-level localization for the repair of all bugs.


\subsubsection{Implementation} 
In our selected datasets, the provided unit tests may not always target the buggy functions; they might instead target the callers of these buggy functions. To obtain test cases specifically designed for the buggy functions, we employ an instrumentation process. Specifically, we use the Java lib-Javassist~\cite{Javassist} to insert instrumentation code at the beginning of the buggy functions to capture the parameter values during the execution of unit tests. These captured values effectively serve as test cases tailored for the buggy functions. Once we have gathered the test cases, we apply our type-aware mutation strategy to the failing tests, resulting in a set of mutated test cases. For executing the mutated test cases, we utilize the original test driver employed by Defects4j. During the execution of the test driver, the instrumented code within the buggy functions substitutes the values of the parameters with the values from the mutated test cases.

\subsubsection{Baselines}
In our comparative evaluation, we evaluate \tool by comparing it with seven state-of-the-art baselines, including six learning-based APR methods (SelfAPR~\cite{ye2022selfapr}, AlphaRepair~\cite{10.1145/3540250.3549101}, RewardRepair~\cite{ye2022neural}, Recoder~\cite{zhu2021syntax}, and CURE~\cite{jiang2021cure}), one traditional APR method (TBar~\cite{liu2019tbar}), and one recent LLM-based technique, CHATREPAIR~\cite{xia2023keep}. 
Additionally, we create an LLM-based baseline called BaseChatGPT, which uses a basic prompt without detailed feedback, such as test pairs and traceback information. BaseChatGPT takes the buggy function as input and iteratively outputs the patched functions. We selected only one traditional APR method because: 1) TBar is the state-of-the-art traditional method and 2) previous research ~\cite{lutellier2020coconut,10.1145/3540250.3549101,xia2023keep,ye2022selfapr,zhu2021syntax} has demonstrated that their performance is usually inferior to learning-based methods.

We chose CHATREPAIR due to two main reasons: firstly, both \tool and CHATREPAIR utilize the state-of-the-art LLM, ChatGPT, making the comparison more relevant. Secondly, although not yet accepted at the time of our submission, CHATREPAIR has achieved new state-of-the-art results, as reported in~\cite{xia2023keep}, surpassing the state-of-the-art methods. Since CHATREPAIR is not publicly open-sourced, so we re-implement it following the instructions provided in the paper~\cite{xia2023keep}. 



\subsubsection{Metrics} We select two widely-used metrics to compare \tool with the baselines:
\begin{itemize}[leftmargin=*]
    \item \textit{Number of Correct Fixes ($\#Correct$)}: Assesses the ability of the repair tool to produce accurate patches. This metric counts the number of programs that have been properly repaired based on a manual review 
    of the plausible patches generated by each tool.
    \item 
\textit{Number of Queries to ChatGPT ($\#Query$)}: Quantifies the resource utilization and efficiency of the LLM-based method. We evaluate the average number of ChatGPT API queries made across all bug cases, providing insight into the frequency with which ChatGPT calls are needed.



\end{itemize}



\begin{table*}[!t]
    \caption{Comparative results with on Defects4j.}
    \resizebox{1\linewidth}{!}{
        \begin{tabular}{lcccccccccc}
        \toprule 
        \textbf{Dataset} & \textbf{\textit{\tool}} & \textbf{BaseChatGPT} & \textbf{CHATREPAIR} & \textbf{AlphaRepair} & \textbf{SelfAPR} & \textbf{RewardRepair*} & \textbf{Recoder} & \textbf{TBar*} & \textbf{CURE} \\
        \midrule 
        \textbf{Chart} & {12} & 10 & 11 & 8 & 9 & 5 & 11 & 11 & 9 \\
        \textbf{Closure} & {32} & 19 & 30 & 22 & 18 & 15 & 22 & 16 & 13\\
        \textbf{Lang} & {19} & 10 & 17 & 11 & 12 & 7 & 9 & 13 & 9\\
        \textbf{Math} & {30} & 23 & 26 & 21 & 15 & 19 & 22 & 22 & 16\\
        \textbf{Mockito} & {8} & 5 & 5 & 5 & 2 & 3 & 1 & 3 & 4\\
        \textbf{Time} & 2 & 1 & 1 & 3 & 3 & 1 & 2 & 3 & 1\\
        \midrule 
        \textbf{D4J1.2} & \textbf{103} & 68 & 90 & 70 & 59 & 50 & 67 & 68 & 52\\ 
        \textbf{D4J2.0} & \textbf{40} & 28 & 34 & 36 & 31 & 25 & 18 & 8 & 19\\
        \bottomrule
        \end{tabular}
    }
    \label{tab:RQ1-defect4j}
\end{table*}

\begin{figure*}[!t]
    \centering
     \includegraphics[width=1\linewidth]{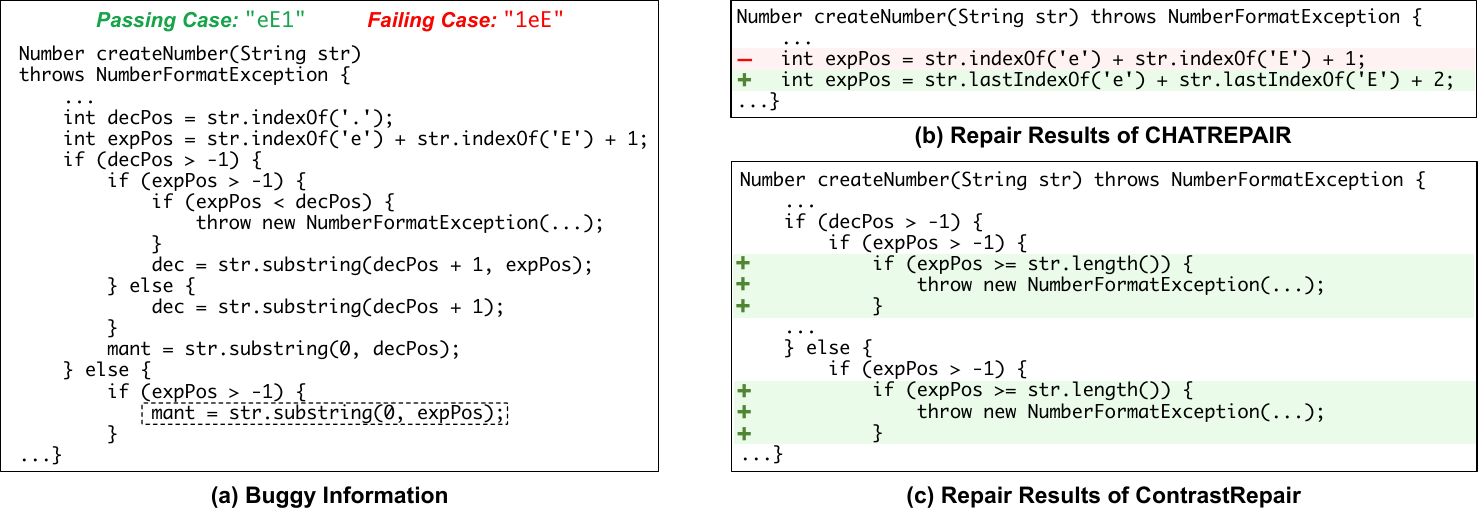}
    \caption{An illustrative example of a bug uniquely fixed by \tool in Defects4J 1.2.}
    \label{fig:CaseStudy}
\end{figure*}

\begin{figure}[!t]
    \centering
        \centering
        \includegraphics[width=0.6\linewidth]{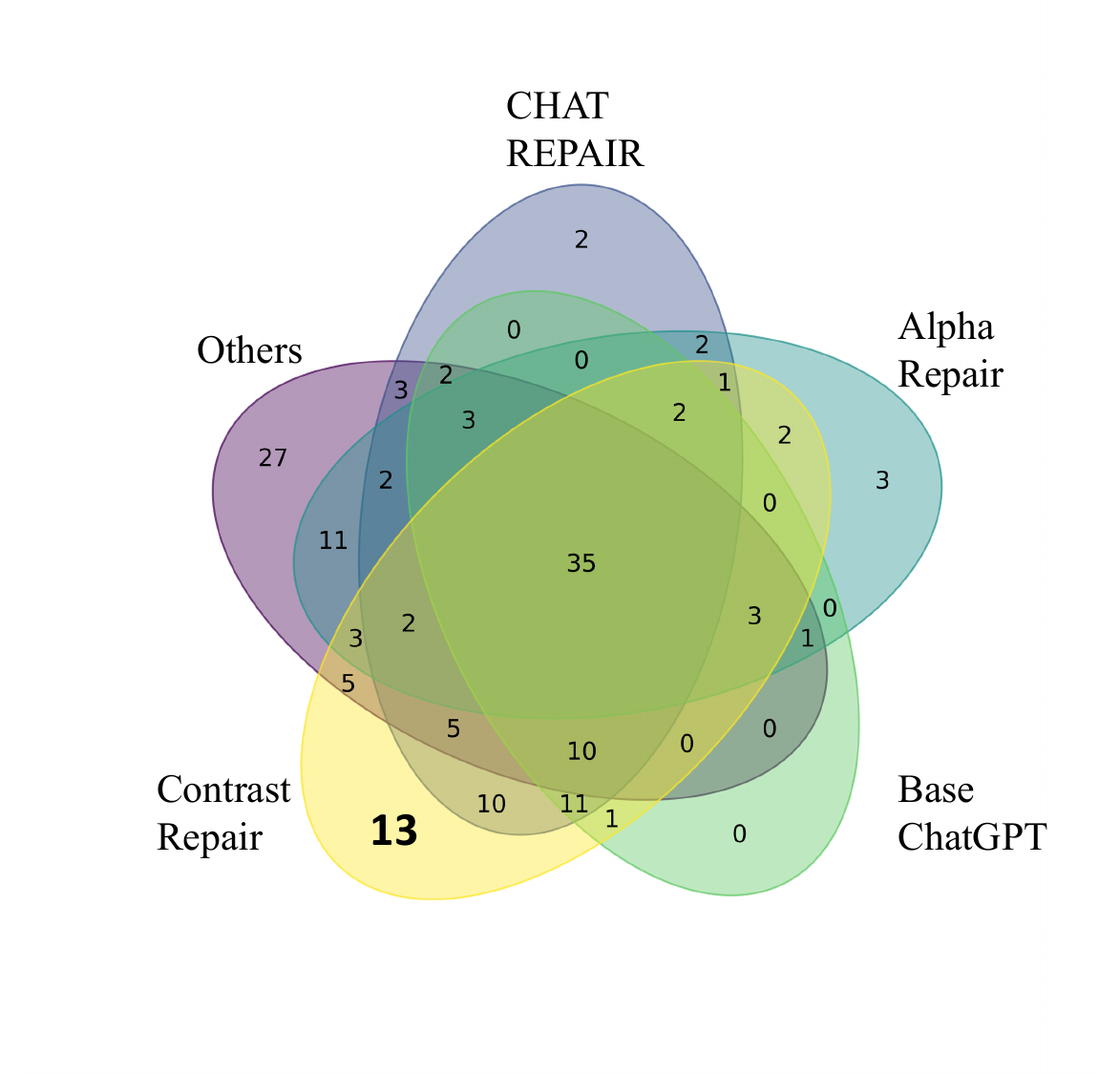}
             \vspace{-8mm} 
        \caption{Bug fix Venn diagram on D4J1.2.}
        \label{fig:vennFig}
    
\end{figure}

\subsection{RQ1: Effectiveness of \tool}

\textbf{Repair Performance on Defects4j.} Table~\ref{tab:RQ1-defect4j} presents the results of different tools on the number of correct fixes for Defects4j 1.2 and Defects4j 2.0 datasets, where the tools are represented by abbreviation. Notably, \tool, BaseChatGPT and CHATREPAIR's results on Defects4j 1.2 encompass the union of single-function fix, single-hunk fix, and single-line fix. The symbol * indicates that the results were not obtained by us, but rather collected from their paper~\cite{xia2023keep} due to the code not being open-sourced or the failure to run in our local configuration. However, the configuration of them are the same as that of the other baselines, and we firmly believe that this does not affect the validity of our comparative conclusions, as \tool consistently demonstrates significant improvements compared to these baselines. As the results suggest, \tool achieves the best performance compared to other baselines. Specifically, \tool can repair 103 out of 255 bugs and 40 out of 82 bugs in Defects4j 1.2 and Defects4j 2.0, respectively. These results set a new state-of-the-art repair performance, showcasing the effectiveness of \tool. 

Comparing the LLM-based tools with other methods, we can clearly observe that integrating ChatGPT significantly improves the repair performance. For instance, the best non-LLM method, AlphaRepair, correctly repairs 70 and 36 bugs on Defects4j 1.2 and 2.0, respectively. On the other hand, BaseChatGPT can achieve competitive performance, i.e., correctly repairs 68 and 28 bugs, demonstrating the usefulness of LLM in the repair task.

Comparing \tool with other LLM-based methods, we can observe that both \tool and CHATREPAIR achieve better results than BaseChatGPT, highlighting the significance of providing informative feedback to LLM in program repair. Note that, there is a gap between the results of CHATREPAIR obtained in this paper and that in the original paper ~\cite{xia2023keep}. The main reason, leading to that the results of CHATREPAIR reported in the original paper are superior, is the variation of hyperparameter settings. In the original paper, the default setting for the maximum number of restarting repair attempts allowed is 200 for single-line and single-hunk scenarios, and 100 for the single-function scenario, which are much larger than ours. A further comparison between \tool and CHATREPAIR is presented in Table~\ref{tab:RQ1-comparisonOnD4j}, where columns SL, SH and SF show the results of single-line fixes, single-hunk fixes and single-function fixes, respectively. As we can see, among all the fix scenarios, \tool performs better than the others. In the three repair scenarios of Defects4j 1.2, \tool significantly outperforms CHATREPAIR by repairing 11, 16, and 23 more bugs respectively. In the single-line-fix scenario of Defects4j 2.0, \tool outperforms CHATREPAIR with 40 correct bug fixes compared to 34. Furthermore, we found that \tool generates about 20\% more plausible fix than CHATREPAIR, 56 versus 47. The improvements demonstrated that \tool could provide more useful prompts, such as the test pairs, which contributes to its superior performance.


\noindent \textbf{Repair Performance on QuixBugs.} Table~\ref{tab:RQ1-quixbugs} displays the results for QuixBugs. Notably, \tool successfully rectified all bugs within the QuixBugs-Java and QuixBugs-Python datasets, demonstrating its marked superiority over other methods. Detailed repair results across three distinct scenarios are presented in Table~\ref{tab:RQ1-quixBugs_3scenario}. 
Specifically, \tool outperforms CHATREPAIR among three scenarios in both QuixBugs-Python and QuixBugs-Java. In QuixBugs-Python, \tool can respectively fix three, one and three more bugs than CHATREPAIR in three different scenarios. Similarly, in QuixBugs-Java, it can perform better by giving two, two and three more correct fixes.

\begin{table}[!t]
    \centering
    \caption{Comparison with baselines on QuixBugs}
    \resizebox{0.7\linewidth}{!}{
        \begin{tabular}{l|ccccc}
        \toprule
        \textbf{QuixBugs} & \makecell[l]{\textit{\textbf{Constrast}}\\ \textit{\textbf{Repair}}} & \makecell[l]{\textbf{CHAT}\\ \textbf{REPAIR}} & \makecell[l]{\textbf{Alpha}\\ \textbf{Repair}} & \makecell[l]{\textbf{Reward}\\ \textbf{Repair}} & \makecell[l]{\textbf{CURE}} \\
        \midrule
        \textbf{Python} & \textbf{40} & 38 & 27 & - & -    \\
        \textbf{Java}   & \textbf{40} & 37 & 28 & 20 & 26  \\
        \bottomrule
        \end{tabular}
    }\label{tab:RQ1-quixbugs}
  \hspace{2pt} 
\end{table}

\begin{table}[!t]
        \centering
        \caption{Comparison on Distinct Scenarios.}
        \resizebox{0.7\linewidth}{!}{
            \begin{tabular}{l|ccccccccc}
            \toprule
            \multirow{2.5}{*}{\textbf{Tools}} & \multicolumn{3}{c}{\textbf{D4j1.2}} & \multicolumn{3}{c}{\textbf{QuixBugs-Py}} & \multicolumn{3}{c}{\textbf{QuixBugs-J}}  \\
             \cmidrule(lr){2-4} \cmidrule(lr){5-7}  \cmidrule(lr){8-10}
             & \textbf{SL} & \textbf{SH} & \textbf{SF} & \textbf{SL} & \textbf{SH} & \textbf{SF} & \textbf{SL} & \textbf{SH} & \textbf{SF}  \\
            \midrule
            \textbf{BaseChatGPT} &38 &50 &37 &31 &32 &30 &29 &30 &33  \\
            \textbf{CHATREPAIR} &45 &53 &52 &33 &35 &34 &32 &33 &34  \\
            \textbf{\tool} &\textbf{56} &\textbf{69} &\textbf{75} &\textbf{36} &\textbf{36} &\textbf{37} &\textbf{34} &\textbf{35} &\textbf{37}   \\
            \bottomrule
            \end{tabular}
        }\label{tab:RQ1-quixBugs_3scenario}
\end{table}

\noindent \textbf{Repair Efficiency.} The repair efficiency in terms of the number of API calls is reported in Table~\ref{tab:RQ1-comparisonOnD4j}. Compared to CHATREPAIR, \tool exhibited an enormous reduction in average query times among all bug cases ($\#Query$), especially in single-line and single-hunk repair scenarios, with $\#Query$ decreased by 28.83\% and 30.20\%, respectively. It indicates that \tool employs better prompts, allowing the LLM to pinpoint bugs and enabling more effective repair. 
In the single-function repair scenario, we noticed that $\#Query$ only decreased by 5.97\%. After analyzing the experiment results, we found that \tool repaired some bugs that CHATREPAIR was unable to fix. Patches of these bugs are more complicated, \tool could only address them when it approached the maximum number of attempts, leading to a marginal decrease in $\#Query$.

\begin{table}[!t]
   \centering
        \caption{Comparison with Baselines on D4J1.2.}
        \resizebox{0.8\linewidth}{!}{
            \begin{tabular}{l|ccccccccc}
            \toprule
            \multirow{2.5}{*}{\textbf{Tools}} & \multicolumn{3}{c}{\textbf{$\#Correct$}} & \multicolumn{3}{c}{\textbf{$\#Plausible$}} & \multicolumn{3}{c}{\textbf{$\#Query$}}  \\
             \cmidrule(lr){2-4} \cmidrule(lr){5-7}  \cmidrule(lr){8-10}
             & \textbf{SL} & \textbf{SH} & \textbf{SF} & \textbf{SL} & \textbf{SH} & \textbf{SF} & \textbf{SL} & \textbf{SH} & \textbf{SF} \\
            \midrule
            \textbf{BaseChatGPT} &38 &50 &37 &43 &67 &67 &66.88 &79.89 &96.16  \\
            \textbf{CHATREPAIR} &45 &53 &52 &47 &70 &101 &62.36 &79.14 &82.27 \\
            \textbf{\tool} &\textbf{56} &\textbf{69} &\textbf{75} &\textbf{60} &\textbf{102} &\textbf{120} &\textbf{44.38} &\textbf{55.24} &\textbf{77.36}   \\
            \bottomrule
            \end{tabular}
        }\label{tab:RQ1-comparisonOnD4j}
\end{table}

Fig~\ref{fig:vennFig} shows a Venn diagram of the bugs fixed by all baselines and \tool on Defects4j 1.2. We selected the top three baselines based on the number of bugs they fixed correctly and grouped the remaining ones under ``Other''. We can see that \tool fixes 13 unique bugs not addressed by any other prior tools. 

Fig~\ref{fig:CaseStudy} presents a specific example (Lang-27) that was only fixed by \tool. This bug manifested as an unexpected  StringIndexOutOfBoundsException error on the highlighted line. In Fig~\ref{fig:CaseStudy} (a), we showcase the buggy function and highlight the line where failing cases trigger the unexpected exception. The input is considered passed if it either parses successfully or raises a NumberFormatException when it is invalid.
As depicted in Fig~\ref{fig:CaseStudy} (b), CHATREPAIR erroneously directs ChatGPT's focus toward the value of \texttt{expPos}, leading to an incorrect adjustment in its calculation. When provided with the failing test case \texttt{1eE}, \tool initially generates a similar passing test case, \texttt{eE1}. 
We conjecture that the passing test could possibly imply that the calculation of \texttt{expPos} is fine and a proper fix can be attained by adding a bounds check.
Additionally, the values of \texttt{expPos} for the passing test \texttt{eE1} and the failing test \texttt{1eE} are \texttt{expPos=2} and \texttt{expPos=4}, respectively, with both inputs having a length of 3. This contrastive information (4>3 and 2<3) might provide stronger guidance for ChatGPT to infer that \textit{the bug is triggered when the value of \texttt{expPos} exceeds the input length}.
As a result, as shown in Fig~\ref{fig:CaseStudy} (c), \tool successfully fixes the bug.

\begin{figure}[!t]
    \centering
    \includegraphics[width=.8\linewidth]{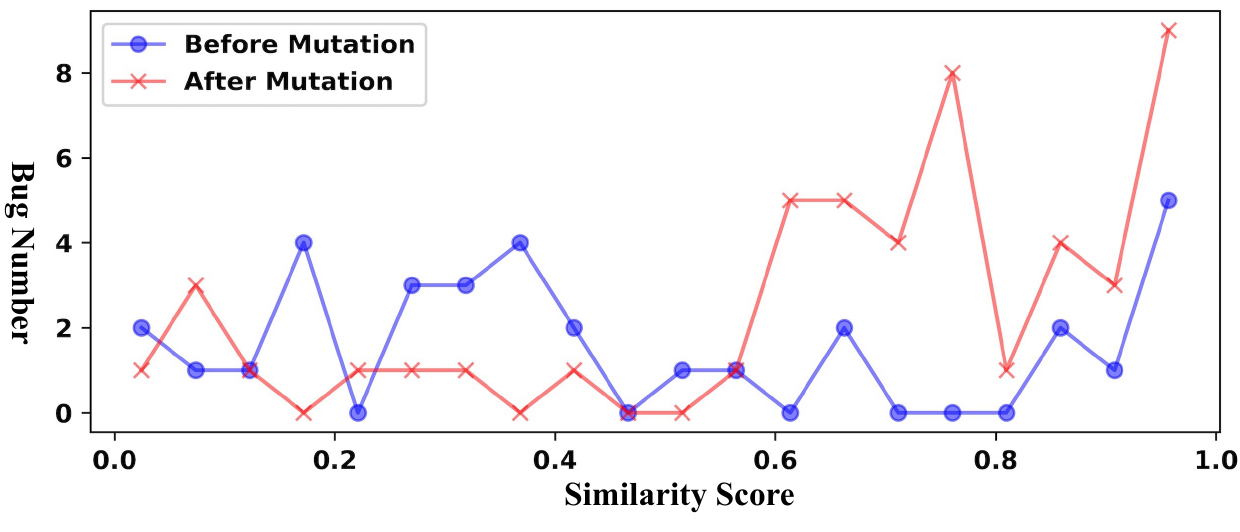}
    \caption{Similar Scores before and after Mutation.}
    \label{fig:Similar_scores}
\end{figure}

\noindent \textbf{Performance of type-aware mutation.} We also evaluate the effectiveness of the proposed type-aware mutation. We find that when applying the mutation operations, we can have 45 extra bugs that can successfully generate passing test cases on the Defects4j Dataset, which demonstrate the effectiveness of type-aware mutation. The remaining bugs we can not successfully generate passing test cases are because these buggy functions have no parameters in primitive type or all parameters are objects instantiated from classes defined by developers. We also analyzed the test pair similarity before and after type-aware mutation. Fig~\ref{fig:Similar_scores} displays the frequency distribution of maximum similarity scores before and after mutation. The x-axis denotes similarity scores, while the y-axis represents the number of bugs. For each bug, we calculate the highest similarity score among its multiple test pairs. Note that a bug receives a maximum similarity score of 0 if it has no test pairs. The original number of this kind of bug is 43, however, after mutation, it decreases to 25. The results reveal a wide distribution of initial similarity scores, with only 14.67\% (11/75) of the bugs falling between 0.5 and 1. However, after mutation, these scores predominantly cluster in the 0.6 to 1 range, encompassing 53.33\% (40/75) of the bugs. This indicates a notable enhancement in similarity scores attributable to our mutation strategy.
For each bug, we randomly generate 1,000 test cases and validate them via the command provided by Defects4j, the time cost of this process ranges from 15 to 25 minutes based on the complexity of different buggy functions.


\noindent \textbf{Answer to RQ1: }
\tool improves the state-of-the-art APR tools significantly, including ChatGPT-based repair method CHATREPAIR. For instance, in the three fix scenario of Defects4j 1.2, \tool achieves a success repair rate of 70.00\%, 44.81\% and 29.41\% respectively, whereas CHATREPAIR achieves 56.25\%, 34.42\% and 20.39\% respectively. Furthermore, \tool requires fewer API queries than CHATREPAIR.

\begin{table}[!t]
      \centering
    \caption{Comparison on HumanEval-Java.}
    \resizebox{0.5\linewidth}{!}{
        \begin{tabular}{l|ccc}
        \toprule
        
        \textbf{Tools} & \textbf{$\#Plausible$} & \textbf{$\#Correct$} &  \textbf{$\#Query$} \\
        \midrule
        \textbf{BaseChatGPT} &136 & 126 & 28.90 \\
        \textbf{CHATREPAIR} &143 & 130 & 24.68\\
        \textbf{\tool} &\textbf{151} & \textbf{137} &\textbf{17.44} \\
        \bottomrule
        \end{tabular}
    }\label{tab:RQ4-humaneval}
\end{table}

\vspace{-2mm}
\subsection{\textbf{RQ2}: Evaluation on Unknown Dataset}
 In light of concerns surrounding potential data leaks from the widely-used benchmarks, Defects4j and QuixBugs, which could be included in the training data of ChatGPT, we conducted experiments to assess the effectiveness of \tool on unseen new datasets. It is worth noting that while there is a possibility of data influence from Defects4j and QuixBugs, our comparative analysis with ChatRepair and BaseChatGPT in RQ1 still underlines the effectiveness of \tool, given that all three methods employ the same version of ChatGPT. To explore a fresh perspective, we opted to utilize the recently introduced benchmark, HumanEval-Java, which was made publicly available in January 2023. In our evaluation, \tool was configured for single-function fixes with setting $n$ and $m$ as 3 and 40, and we compared the outcomes with those obtained using CHATREPAIR and BaseChatGPT. 

As illustrated in Table~\ref{tab:RQ4-humaneval}, \tool achieves the best results among the three tools, successfully fixing 137 out of 164 total bugs. In comparison, CHATREPAIR achieved 130 fixes, while BaseChatGPT managed 126. Notably, \tool exhibited an improvement in the success repair rate by 5.38\% and 8.73\% when compared to CHATREPAIR and BaseChatGPT, respectively. 

Furthermore, \tool significantly outperforms the other two methods in terms of efficiency, as indicated by the average number of ChatGPT queries on each bug ($\#Query$). Specifically, \tool reduced the number of $\#Query$ by 29.34\% (from 24.68 to 17.44) in comparison to CHATREPAIR, and by 39.66\% (from 28.90 to 17.44) when compared to BaseChatGPT. These results are consistent with our findings from the Defects4j benchmark, demonstrating its effectiveness.

\noindent \textbf{Answer to RQ2: }
The evaluation of the unseen dataset, HumanEval-Java, again showcases the superior performance of  \tool in terms of both the number of correct fixes and the efficiency of ChatGPT API queries.

\begin{table}[!t]
    \centering
    \caption{Evaluation on Pair Number.}
    \resizebox{0.7\linewidth}{!}{
        \begin{tabular}{l|cc|cccc}
        \toprule
        
        \textbf{Settings} & \textbf{\#Fail} & \textbf{\#Pair} & \textbf{$\#Correct$} & \textbf{$\#Plausible$} & \textbf{$\#Query$} \\
        \midrule
        \textbf{CHATREPAIR} &0 &0 &33 &54 &69.09\\ \hline
        \textbf{BaseChatGPT} &0 &0 &31 &48 &76.16\\
        \textbf{+SingleFail} & 1 & 0 &32 &53 &69.23\\
        \textbf{+DoubleFail} & 2 & 0 &33 &55 &68.57\\
        \textbf{+SinglePair} & 0 & 1 &35 &58 &66.14\\
        \textbf{\tool} & 0 & 2 &\textbf{37} &\textbf{60} &\textbf{63.09}\\
        \bottomrule
        \end{tabular}
    }
      \label{tab:RQ2-pairNumber}
\end{table}

\begin{table}[!t]
    \centering
    \caption{Evaluation on Restarting/Continuous Repair.}
    \resizebox{.7\linewidth}{!}{
        \begin{tabular}{cc|cccc}
        \toprule
        \textbf{Iteration ($m$)} & \textbf{Rounds ($n$)} & \textbf{$\#Correct$} & \textbf{$\#Plausible$} & \textbf{$\#Query$} \\
        \midrule
        12 & 10 &15 &23 &88.48 \\
        120 & 1 &18 &31 &74.78 \\
        40 & 3 &\textbf{21} &\textbf{33} &\textbf{70.53} \\
        \bottomrule
        \end{tabular}
        \label{tab:RQ2-mxn}
    }
\end{table}

\subsection{RQ3: Hyperparameter Evaluation} \label{rq3}

The primary hyperparameters for \tool encompass the number of test case pairs ($k$) and the total number of attempts, determined by the continuous and restarting repair configuration ($m, n$). We perform the single-function repair as it is a more practical scenario, and the dataset consists of 106 bugs (from Defects4j) in which at least one pair can be constructed.
Table~\ref{tab:RQ2-pairNumber} shows the outcomes from various test case configurations (Columns \textit{\#Fail} and \textit{\#Pair}). These configurations span four variants: a single failed case, two failed cases, one test pair, and two test pairs. We present the repair results of LLM-based baselines on these 106 bugs. We limit the maximum number of test pairs to 2, as half of the bugs in the dataset have fewer than 2 pairs of test cases with a similarity threshold of 0.5. For consistency, we only retrain using the top-2 test pairs when a bug has more than two pairs.
Comparisons were made based on the correct fixes (Column \textit{\#Correct}), plausible fixes (Column \textit{\#Plausible}), ChatGPT API query count (Column \textit{$\#Query$}).  

Our observations revealed that \tool, when provided with two test case pairs, yielded the highest correct and plausible fix rates (37/60). Offering just a single test case pair led to the second-highest fix effectiveness (35/58). Crucially, using test case pairs consistently outperformed merely submitting the same quantity of failed test cases. Moreover, introducing two failed test cases (33/55) was more effective than just one (32/53). Notably, \tool, when fed with two test case pairs, demanded the fewest $\#Query$ (63.09). These findings underscore the value of test pairs: they deliver more useful information to LLMs for bug repair than simply relaying failed tests. Consequently, supplying an increased number of tests or test pairs tends to optimize results.

For our restarting repair and continuous repair parameters ($m \times n$), we considered three distinct configurations: $12 \times 10$, $40 \times 3$, and $120 \times 1$. Each configuration adheres to the same overall budget limit, totaling 120 tries. However, they differ in their approach: the $12 \times 10$ setup emphasizes a deeper search, the $120 \times 1$ aims for a broader search, while $40 \times 3$ strikes a balance between the two.

An analysis of the outcomes in Table~\ref{tab:RQ2-mxn} reveals that the $40 \times 3$ configuration excels in terms of the number of correct/plausible fixes (21/33). Conversely, $12 \times 10$ exhibits the least effectiveness (15/23). This disparity suggests that the parameter $n$ is pivotal in influencing \tool's performance.
A larger $n$ seems detrimental: if an initial repair attempt deviates from the optimal solution, realigning to a correct path becomes challenging as iterations increase. This can hinder the generation of accurate patches. However, setting $n$ too low can be counterproductive. If the initial direction of repair is on track but achieving the desired patch demands successive iterations, a moderately large $n$ becomes beneficial. Continuous feedback in such cases can aid in formulating the correct fixes.

\noindent \textbf{Answer to RQ3: }
Utilizing test pairs provides more comprehensive information to LLMs for effective and efficient bug repair compared to solely relying on failed tests. The balance between restarting and continuous repairs is crucial to ensure the repair performance.



\begin{table}[!t]
    \centering
    \caption{Effectiveness of Selection and Contextual Code.}
    \resizebox{0.6\linewidth}{!}{
        \begin{tabular}{l|ccc}
        \toprule
        \textbf{Tools} & \textbf{$\#Correct$} & \textbf{$\#Plausible$} &  \textbf{$\#Query$}   \\
        \midrule
        \textbf{w/o Pair} &30.67 &48.00 &76.16\\
        \textbf{w/o Similarity} &33.33 &56.33 &72.62\\
        \textbf{\tool-106} &\textbf{37.00} &\textbf{60.33} &\textbf{63.09}\\
        \hline
        \textbf{w/o Context} & 32.67 & 61.67  &62.17 \\
        \textbf{\tool-103} & \textbf{36.00} & \textbf{64.67} & \textbf{58.28} \\
        \bottomrule
        \end{tabular}
    }
    \label{tab:RQ4-selection}
\end{table}

\subsection{RQ4: Ablation Study on \tool}

{In this section, we delve into an ablation study to evaluate the usefulness of two critical components in \tool: pair selection and dependent functions. Specifically, we configure three variants of our tool: \tool without test pairs (referred to as \textbf{w/o Pair}), the tool that replaces the similarity-based selection with random selection (referred to as \textbf{w/o Similarity}) and \tool  without dependent functions (referred to as \textbf{w/o Context}).}

To evaluate the evaluation of \textbf{w/o Pair} and \textbf{w/o Similarity}, we employ the same dataset used in RQ3, consisting of 106 bugs, each with multiple test pairs. For \textbf{w/o Context}, we chose all the 103 bugs from Defects4j where each sample has multiple dependent functions. The reason for this selection is that not each sample has the function invocation or test pair. Similar to RQ3, we repair the bugs in a practical setting, i.e., single-function fix. Moreover, to mitigate the effect of randomness, we repeated the bug rectification process three times for each bug.



{As indicated in Table~\ref{tab:RQ4-selection}, the results underscore the importance of test pair, dependent functions and the similarity-based selection used in \tool.} For the 106 bugs, \tool generates 60.33 plausible fixes and 37.00 correct fixes on average. 
\tool \textbf{w/o Pair} only has 48.00 plausible fixes and 30.67 correct fixes on average. However, when we replace the similarity selection with the random selection that may not select the pair with high similarity, the results are 56.33 plausible fixes and 34.00 correct fixes on average, which is higher than the setting without pairs but lower than \tool. This highlights an enhancement achieved by including the similarity-based pair selection.
For the 103 bugs where \tool leverages the contextual information regarding dependent functions, it produces 64.67 plausible fixes and 36.00 correct fixes on average. Conversely, {\textbf{w/o Context}}, without the contextual information, generates 61.67 plausible fixes and 32.67 correct fixes. 

To demonstrate the value of each component, we adopted an additional metric, Pass@m, with $m$ indicating the maximum number of repair attempts allowed (refer to Algorithm 1). We set varying values for $m$, as detailed in Table~\ref{tab:RQ4-pass_k_similarity}.
In this table, \textit{ours} denotes the results achieved by \tool on various datasets, which contained 106 and 103 bugs, respectively.
Overall, the results continue to affirm the usefulness of test pairs, similarity-based test pair selection, and the context involving dependent functions. Notably, with a smaller $m$ (i.e., a tighter budget), the impact of these components on generating correct ($\#C$) and plausible ($\#P$) patches becomes more pronounced, as seen in Pass@20, Pass@10, and Pass@5.
Furthermore, the incorporation of {similarity-based pair selection} and contextual information contributes to a reduction in the number of queries made to ChatGPT API. For instance, on the two selected datasets, the average number of queries per bug for \tool is 63.09 and 58.28 respectively. {In contrast, \textbf{w/o Pair}, \textbf{w/o Similarity} and \textbf{w/o Context} have averages of 76.16,  72.62, and 62.17, representing increases of 20.72\%, 15.11\% and 6.67\%, respectively.

\noindent \textbf{Answer to RQ4: }
Incorporating {similarity-based pair selection} and dependent functions in prompts helps enhance the capabilities of \tool for bug repair. These components not only significantly improve the effectiveness of LLMs in generating more correct fixes (8.82\% and 9.09\%) but also reduce the number of API calls ({20.72\%, 15.11\% and 6.67\%}). 

\begin{table}[!t]
    \caption{Pass@m for ablation on Similarity and Context.}
    \resizebox{\linewidth}{!}{
        \begin{tabular}{l|cccccccccc}
        \hline 
        \multirow{2}{*}{\textbf{Tools}}& \multicolumn{2}{c}{\textbf{Pass@40}} & \multicolumn{2}{c}{\textbf{Pass@30}} & \multicolumn{2}{c}{\textbf{Pass@20}} & \multicolumn{2}{c}{\textbf{Pass@10}} & \multicolumn{2}{c}{\textbf{Pass@5}}\tabularnewline
         & $\#C$ & $\#P$ & $\#C$ & $\#P$ & $\#C$ & $\#P$ & $\#C$ & $\#P$ & $\#C$ & $\#P$\tabularnewline \hline 
        \textbf{w/o Pair} & 30.67 & 48.00 & 29.67 & 45.00 & 28.33 & 41.00 & 24.00 & 37.33 & 22.67 & 36.00\tabularnewline
        \textbf{w/o Sim.} & 33.33 & 56.33 & 32.00 & 51.33 & 27.67 & 43.67 & 23.33 & 37.67 & 22.33 & 32.00\tabularnewline
        \textbf{\tool (106)} & \textbf{37.00} & \textbf{60.33} & \textbf{36.00} & \textbf{59.00} & \textbf{33.67} & \textbf{55.33} & \textbf{31.67} & \textbf{50.00} & \textbf{28.67} & \textbf{42.33}\tabularnewline
        \hline 
        \textbf{w/o Con.} & 32.67 & 61.67 & 32.00 & 59.67 & 29.00 & 54.33 & 24.67 & 46.00 & 20.00 & 31.33\tabularnewline
        \textbf{\tool (103)} & \textbf{36.00} & \textbf{64.67} & \textbf{34.67} & \textbf{61.33} & \textbf{33.00} & \textbf{58.67} & \textbf{30.33} & \textbf{52.00} & \textbf{25.33} & \textbf{41.33}\tabularnewline
        \hline
        \end{tabular}
    }
    \label{tab:RQ4-pass_k_similarity}
\end{table}

\section{Discussion}
\subsection{Selection with Different Similarity Strategies}
To investigate the effectiveness of our selection strategy, we use BM25~\cite{robertson2009probabilistic} (lexical-based) and CodeBERT~\cite{feng2020codebert}, UniXcoder~\cite{guo2022unixcoder} (semantic-based) approaches for comparison. Compared with our used \textit{Damerau-Levenshtein distance} for similarity measurement, BM25 utilized sparse vector representation for lexical matching. CodeBERT and UniXcoder are both learn-based approaches where CodeBERT learns code representations by the masked language modeling from a large code corpus and UniXcoder further incorporates AST with different pre-training tasks for learning. Like the configuration in RQ4, we explore the effectiveness of the 106 bugs with multiple pairs. The experiments are repeated three times to avoid randomness. Furthermore, for CodeBERT and UniXcoder, we feed the test cases to the model and obtain the vector representation to calculate the similarity based on the cosine function. 

The experimental results are presented in Table~\ref{tab:Discussion_similarity}, and we can find that using \textit{Damerau-Levenshtein distance} is superior to other approaches. As the objective of similarity measurement is to find the critical changes in the test cases, hence using the vector representation for retrieval may be indirect and inappropriate for this task rather than string-based similarity retrieval. The experimental results in Table~\ref{tab:Discussion_similarity} have demonstrated the effectiveness of our similarity measurement approach,

\begin{table}[!t]
    \caption{Pass@m to investigate the effectiveness of different similarity strategies to select cases.}
    \resizebox{\linewidth}{!}{
        \begin{tabular}{l|cccccccccc}
        \hline 
        \multirow{2}{*}{\textbf{Tools}}& \multicolumn{2}{c}{\textbf{Pass@40}} & \multicolumn{2}{c}{\textbf{Pass@30}} & \multicolumn{2}{c}{\textbf{Pass@20}} & \multicolumn{2}{c}{\textbf{Pass@10}} & \multicolumn{2}{c}{\textbf{Pass@5}}\tabularnewline
         & $\#C$ & $\#P$ & $\#C$ & $\#P$ & $\#C$ & $\#P$ & $\#C$ & $\#P$ & $\#C$ & $\#P$\tabularnewline \hline 
        \textbf{BM25} & 35.67 & 59.00 & 33.00 & 55.33 & 33.33 & 53.33 & 27.33 & 45.67 & 24.67 & 38.00\tabularnewline
        \textbf{CodeBERT} & 33.00 & 57.33 & 32.67 & 53.00 & 29.00 & 51.00 & 27.67 & 45.00 & 26.33 & 38.00\tabularnewline
        \textbf{UniXcoder} & 32.67 & 54.00 & 31.33 & 51.33 & 28.00 & 45.33 & 27.33 & 39.67 & 26.33 & 36.00\tabularnewline
        \textbf{\tool (106)} & \textbf{37.00} & \textbf{60.33} & \textbf{36.00} & \textbf{59.00} & \textbf{33.67} & \textbf{55.33} & \textbf{31.67} & \textbf{50.00} & \textbf{28.67} & \textbf{42.33}\tabularnewline
        \hline
        \end{tabular}
    }
    \label{tab:Discussion_similarity}
\end{table}

\subsection{Threats to Validity}
The first potential threat to the validity of our results is the selection of the datasets used for evaluation. The improvement obtained by \tool may not generalize to other repair datasets. To mitigate this threat, we have selected diverse datasets for evaluation, including Defects4j 1.2, Defects4j 2.0, QuixBugs. The second threat is data leakage which arises when the ground truth patches are part of the original training data of ChatGPT. To address this, we selected an additional dataset, HumanEval-Java, which was published after the training of the ChatGPT models.



Manual verification of the plausible patches is a threat. 
A careful examination is needed to determine whether they are semantically equivalent. To mitigate the threat, we invited three researchers in SE field to check respectively, and then discuss the patches where validation answers were inconsistent, ultimately reaching a consensus. Additionally, we have made our patches open-source for public evaluation.

The assessment of efficiency in our experiments could potentially present challenges. Notably, we did not measure the runtime of different methods due to two reasons. The main one is we have observed that LLM-based repair methods not only consume significantly less time but also perform much better compared to learning-based repair methods during the experimental process in RQ1. Even if \tool consumes about twice total time (consists of query and mutation phases) as much as ChatRepair does (only contains query), it still remains entirely within an acceptable range. At this juncture, we think prioritizing repair effectiveness more than efficiency could be more reasonable when considering a trade-off. Another one is the presence of numerous uncontrollable factors when invoking ChatGPT API calls, such as network delays, the configuration of ChatGPT models, and their parallelism and scheduling strategies. To mitigate this, we metered the number of API calls (i.e., $\#Query$) to gauge the informativeness of prompts, which helps measure whether ChatGPT better understands the root cause of the bugs. 
Our observations suggest that the performance of conversation-based repair typically improves with an increased number of API calls. By comparing the query counts, we can assess the effectiveness of different feedback in guiding the repair process.
Furthermore, we did not use the total number of tokens as a measure of cost or efficiency. There were two primary reasons: 1) the prompts used in each method still have substantial potential for streamlining and optimization by adjusting their content, which is not our main focus. 2) Our reproduced CHATREPAIR may not precisely match the number of tokens in their original implementation, potentially leading to unfair comparisons. 
To this end, we made a trade-off and selected the number of API calls as a suitable metric for the efficiency assessments.

Another potential threat to the validity of our results is that some of the results from RewardRepair and TBar are collected from existing paper~\cite{xia2023keep}. However, we believe that these comparisons do not affect our conclusions, as there is a significant performance gap~\cite{xia2023keep} between these methods and the LLM-based methods. 


\section{Related Work}
\subsection{Automatic Program Repair}
Automated Program Repair (APR) tools are employed to create patched code from the original code and the corresponding buggy location. Every patch generated by the APR tool undergoes validation against the test suite. Plausible patches refer to those that successfully pass the entire suite, while correct patches are plausible patches that effectively rectify the underlying bug. Generally speaking, there are two kinds of methods in the APR field, traditional and learning-based methods. 

Traditional tools can be broadly divided into three main categories: heuristic-based~\cite{le2016history, le2011genprog, wen2018context}, constraint-based~\cite{le2017s3, long2015staged, mechtaev2016angelix} and template-based~\cite{martinez2016astor, liu2019tbar, liu2019avatar}. However, these traditional methods have some limitations. Template-based tools have been regarded as state-of-the-art ones among traditional methods due to their best repair performance, but these APR tools depend much on manually designed templates or specific fix patterns to repair limited kinds of bugs.

With the rapid development of deep learning techniques, there has been a growing focus on learning-based approaches, such as CURE~\cite{jiang2021cure}, RewardRepair~\cite{ye2022neural}, Recoder~\cite{zhu2021syntax}, CoCoNut~\cite{lutellier2020coconut} and SelfAPR~\cite{ye2022selfapr}, which convert APR to Neural Machine Translation(NMT) problem and have shown remarkable potential for enhancing bug repair performance. Unlike conventional methods, learning-based approaches can automatically capture semantic relations among parallel bug-fixing pairs. This capability enables the creation of patch solutions that are not only more effective but also contextually aware.
Nevertheless, the quality and quantity of the training datasets largely determine the effectiveness of the model. The training corpus may include a lot of noise when scraping repositories that have irrelevant commits and changes from GitHub. 

Recently, researchers have explored the feasibility of employing potent LLMs for APR. LLMs demonstrate the ability to produce accurate patches directly from the contextual information, without the necessity for fine-tuning. AlphaRepair~\cite{10.1145/3540250.3549101}, the first cloze-style APR approach to directly leveraging large pre-trained code models for APR without any fine-tuning/retraining on historical bug fixes.

Despite LLM-based methods have yielded remarkable results~\cite{xia2023automated, prenner2022can, jiang2023impact, sobania2023analysis, 10.1145/3540250.3549101, fan2023automated}, they mainly focus on the buggy code and treat bug repair as a one-step process, overlooking the interactive and collaborative aspects inherent in bug resolution. Furthermore, by mining and analyzing historical data from interaction records with LLMs, and understanding the fundamental reasons for LLM failure in bug fixing, researchers can better guide LLMs to perform program repair in the correct direction in subsequent interactions.

Previous research~\cite{gao2019crash, yang2017better} has also demonstrated the effectiveness of applying test cases to code repair tasks, aiding APR tools in generating patches of high quality. Based on these intuitions, CHATREPAIR~\cite{xia2023keep}, the first fully automated conversation-driven APR approach that interleaves patch generation with instant feedback from test suites to perform APR in a conversational style, is proposed. Compared to traditional and learning-based methods, CHATREPAIR can achieve promising results through the use of LLMs by providing error feedback information for the conversations. Regrettably, this feedback may not consistently provide the precise and informative prompts required for efficient repairs. This paper introduces a method, \tool, which aims to craft more specific and directive prompts for enhancing the capabilities of LLMs in accurately comprehending buggy-related semantics and generating high-quality patches.


\subsection{Large Language Models}
Recent advancements in generative AI have resulted in a significant increase in performance and the widespread embrace of Large Language Models (LLMs)~\cite{zhao2023survey, chang2023survey}. In the Natural Language Processing (NLP) domain, LLMs have achieved impressive performance in many tasks such as machine translation~\cite{sutskever2014sequence}, text summarization~\cite{liu2019fine} and classification~\cite{yang2019xlnet}. Initially excelling in Natural Language Processing (NLP) tasks~\cite{min2023recent}, such as document classification~\cite{hegselmann2023tabllm}, text summarization~\cite{yang2023exploring}, and machine translation~\cite{zhang2023prompting}. Tremendous progress has been made by treating programming language as another sort of natural language and training LLMs on corpora of both program code and natural language text, and LLMs are now widely applied in various software engineering tasks including code generation~\cite{zeng2022extensive,liu2023your}, and code summarization~\cite{ahmed2022few, liu2020retrieval}. Since LLMs are designed to be general and capable of acquiring knowledge across various domains, researchers can then make use of LLMs for related downstream tasks by offering customized prompts or, if desired, a few task-solving demonstrations as input~\cite{liu2023pre}. 

LLMs are advanced language models with massive parameter sizes and exceptional learning capabilities. The core module behind many LLMs such as GPT-3~\cite{floridi2020gpt}, InstructGPT~\cite{ouyang2022training}, and GPT-4~\cite{gpt-4} is the self-attention module in Transformer~\cite{vaswani2017attention} that serves as the fundamental building block for language modeling tasks. An important characteristic of LLMs is their ability to perform in-context learning~\cite{brown2020language}, wherein the model is trained to produce text in accordance with a provided context or prompt. This capability allows LLMs to produce responses that are more coherent and contextually appropriate, rendering them well-suited for interactive and conversational purposes.

Reinforcement Learning from Human Feedback(RLHF)~\cite{christiano2017deep, ziegler2019fine} represents another pivotal element of LLMs. This approach entails fine-tuning the model by utilizing human-generated responses as rewards, enabling the model to learn from mistakes and enhance its performance progressively. Significantly, ChatGPT~\cite{chatgpt}, with its emphasis on conversations and its capability to remember and refer back to previous dialogues, has attained state-of-the-art performance in diverse SE tasks ~\cite{dong2023self, sobania2023analysis}. 
In this paper, our purpose is focusing on how to design more powerful prompts by harnessing and utilizing information from test cases to prompt and inspire ChatGPT to understand the semantic aspects of bugs, achieving more effective bug repair, during the process of interacting and collaborating.


\subsection{Automated Test Generation} Automated test generation is a widely used technique for detecting software defects by generating tests automatically. Techniques like fuzz testing~\cite{miller1990empirical, yang2011finding, holler2012fuzzing} in black-box test generation, involve supplying the system under test (SUT) with random test inputs (e.g., random bytes) without examining its internal code structure. Traditional black-box techniques can mainly be categorized into generation-based~\cite{yang2011finding, holler2012fuzzing} and mutation-based~\cite{winterer2020unusual, cha2015program, oehlert2005violating} ones. Generation-based techniques focus on creating test inputs from scratch, whereas mutation-based approaches introduce systematic alterations to existing seed inputs to craft a wider array of tests. A core drawback, often termed as "blindness," of black-box strategies is their struggle to develop intricately designed test cases that thoroughly investigate complex code pathways. Generation-based techniques focus on directly generating test inputs from scratch, whereas mutation-based techniques apply systematic mutation changes on seed inputs to generate more diverse tests.

As a fundamental limitation of “blindness”, black-box methods may hardly produce well-designed test-cases to exhaustively explore deep code paths. White-box methods yield higher-quality test cases by examining the source code of SUT. For example, symbolic execution~\cite{king1976symbolic, cadar2008klee} overcomes coverage limitations by employing symbolic path constraint to generate tests that target deeper paths. However, due to the high cost of constraint solving, some solutions to enhance scalability have been introduced, including algorithmic approaches like concolic testing~\cite{majumdar2007hybrid, sen2005cute} and implementation strategies such as optimized compilation ~\cite{poeplau2020symbolic}. Serving as an intermediary between black- and white-box testing, coverage-guided fuzzing~\cite{serebryany2016continuous, afl}, also known as grey-box testing, leverages coverage information from SUT as feedback to adjust the process of generating and mutating inputs.

Due to our intention not to increase code coverage or the diversity of test cases, we continue to follow black-box test generation. Inspired by type-aware operator mutation~\cite{winterer2020unusual}, we implement mutations on parameters, which is referred to as type-aware parameter mutation. This strategy serves two purposes: firstly, it enables the rapid generation of a substantial number of test cases in a short period. Secondly, by employing fine-grained mutations that make minimal alterations to the original cases, we ensure that the newly generated cases are highly similar to their predecessors.

\subsection{Test Case Similarity} 
To select similar pairs of test cases, we require certain metrics to measure the similarity of each pair. There are two widely used metrics: 
\textbf{Lexical-based Similarity} and \textbf{Semantic-based Similarity}. BM25~\cite{robertson2009probabilistic} is a lexical-based similarity, which uses sparse vector representation for lexical matching. BM25 converts each code snippet as a bag-of-words representation and computes a lexical similarity between a pair ${<F_i, P_j>}$. The computed similarity score is represented as $f_{\phi}(F_i, P_j) = \text{BM25}(F_i, P_j)$. As a sparse term-based metric, BM25 is sensitive to the choice of identifier naming in source code which does not impact the code semantics. Dense Passage Retriever (DPR)~\cite{karpukhin2020dense} belongs to semantic-based similarity. For training DPR to effectively learn code embeddings, both positive pairs and negative pairs are needed, and this is based on the assumption that the two similar code snippets often share similar semantics (e.g., identifiers and code structures). 
However, in this paper, our test cases are not represented in the form of code snippets, they consist of specific values for the parameters within the buggy function instead. Consequently, the two aforementioned metrics used to assess code similarity are not applicable to our context. Our method involves first converting these values into strings, and then calculating their similarity using the Damerau-Levenshtein distance~\cite{damerau1964technique, levenshtein1966binary}

\section{Conclusion}
In this paper, we introduced \tool, a novel conversation-based APR method that utilizes ChatGPT for repairing bugs. By generating contrastive test pairs, \tool provides informative feedback to ChatGPT, enabling better localization of bug causes. We conducted extensive evaluations on diverse benchmark datasets, including Defects4j, QuixBugs, and HumanEval, and compared \tool with state-of-the-art APR baselines. The results demonstrate that \tool achieves a new state-of-the-art in automated program repair.


\bibliographystyle{ACM-Reference-Format}
\bibliography{reference}


\end{document}